\documentclass[a4paper,twocolumn]{article}
\usepackage{abstract}
\usepackage{amsmath}
\usepackage{authblk}
\usepackage{booktabs}
\usepackage{ccaption}
\usepackage{color}
\usepackage[width=18cm]{geometry}
\usepackage{graphicx}
\usepackage{multirow}
\usepackage{url}
\usepackage[pdfborder={0 0 0},colorlinks=true,citecolor=red,linkcolor=blue]{hyperref}

\newcommand{\mytitle}{Multiplication of sparse Laurent polynomials and Poisson series on modern hardware architectures}

\captiondelim{. }
\captionnamefont{\sc\bfseries}
\captiontitlefont{\itshape}

\title{\mytitle}

\author{Francesco Biscani\footnote{\href{mailto::bluescarni@gmail.com}{bluescarni@gmail.com}}}
\affil{European Space Agency -- Advanced Concepts Team \\ European Space Research and Technology Centre (ESTEC)}

\begin{document}

\maketitle

\begin{abstract}
In this paper we present two algorithms for the multiplication of sparse Laurent polynomials and Poisson series (the latter being algebraic structures commonly arising in
Celestial Mechanics from the application of perturbation theories).
Both algorithms first employ the Kronecker substitution technique to reduce multivariate multiplication to univariate multiplication,
and then use the schoolbook method to perform the univariate multiplication.
The first algorithm, suitable for moderately-sparse multiplication, uses the exponents of the monomials resulting from the univariate
multiplication as trivial hash values in a one dimensional lookup array of coefficients. The second algorithm, suitable for highly-sparse multiplication, uses a cache-optimised hash table
which stores the coefficient-exponent pairs resulting from the multiplication using the exponents as keys.

Both algorithms have been implemented with attention to modern computer hardware architectures. Particular care has been devoted to the efficient exploitation of contemporary
memory hierarchies through cache-blocking techniques and cache-friendly term ordering. The first algorithm has been parallelised for shared-memory multicore architectures, whereas
the second algorithm is in the process of being parallelised.

We present benchmarks comparing our algorithms to the routines of other computer algebra systems, both in sequential and parallel mode.
\end{abstract}


\section{Introduction}
The application of perturbation theories for non-linear differential equations in Celestial Mechanics is a task traditionally associated with cumbersome yet simple operations
on basic algebraic structures. Modern applications of perturbative methods to the problem of the long-term stability of the Solar System, for instance, lead to series expansions
whose number of terms is in the order of $10^5 - 10^6$ \cite{kuz2004,giorg2009}. It is then not surprising that astronomers and celestial mechanicians have sought -- since the dawn of the digital age --
to fully exploit the power and reliability of computers in this domain, producing a vast amount of literature on specialised (as opposed to general-purpose) algebraic manipulators
\cite{musen1959,Broucke69,Jefferys70,Jefferys72,Rom70,CAMAL,UPP,Dasenbrock82,Parsec89,Abad94,Ivanova96,gastineau2006}.

Typically, perturbative methods in Celestial Mechanics require the ability to manipulate symbolically algebraic structures known as Poisson series \cite{San-Juan01}, consisting of Fourier
series having Laurent polynomials as coefficients:
\begin{multline}
\sum_{j_1\ldots j_m}\sum_{i_1\ldots i_n}C_{i_1\ldots i_n,j_1\ldots j_m}x_1^{i_1}\cdots x_n^{i_n} \\ \begin{matrix}
\cos \\ \sin \end{matrix}
\left(j_1y_1+\ldots+j_my_m \right).\label{eq:ps_def}
\end{multline}
Here the symbolic variables are denoted by $x$ and $y$, while $i$ and $j$ are integer indices and $C$ is a numerical coefficient. The mathematical operations to be performed on
such objects are usually elementary. E.g., the computation of normal forms using the Lie series technique, a popular methodology in Celestial Mechanics,
requires the operations of addition, subtraction, multiplication, differentiation and integration \cite{deprit69}.

The most taxing operation to be performed on Poisson series, which also forms the basis for more complicated operations such as Taylor expansions, is series multiplication.
In this paper we present two algorithms for the multiplication of sparse Laurent polynomials and Poisson series which employ the technique of Kronecker substitution and whose implementation
seeks to maximise performance on modern computer hardware architectures.

\section{Kronecker substitution}
Kronecker substitution, first described in \cite{kronecker82},
is a methodology that allows to reduce multivariate polynomial multiplication to univariate multiplication, and it can be intuitively
understood with the aid of a simple table laying out the monomials
in reverse lexicographic order. The representation for three
variables $x$, $y$ and $z$, and up to the third power in each variable
is displayed in Table \ref{tab:Kronecker-substitution-for}.%
\begin{table}
\begin{center}
\begin{tabular}{cccc}
\toprule$z$ & $y$ & $x$ & Code\tabularnewline
\midrule0 & 0 & 0 & 0\tabularnewline
0 & 0 & 1 & 1\tabularnewline
0 & 0 & 2 & 2\tabularnewline
0 & 0 & 3 & 3\tabularnewline
\midrule0 & 1 & 0 & 4\tabularnewline
0 & 1 & 1 & 5\tabularnewline
0 & 1 & 2 & 6\tabularnewline
0 & 1 & 3 & 7\tabularnewline
\midrule0 & 2 & 0 & 8\tabularnewline
$\ldots$ & $\ldots$ & $\ldots$ & $\ldots$\tabularnewline
3 & 3 & 3 & 63\\ \bottomrule\tabularnewline
\end{tabular}
\end{center}

\caption{Kronecker substitution for a 3-variate polynomial up to the third
power in each variable.\label{tab:Kronecker-substitution-for}}

\end{table}
The column of codes is obtained by a simple enumeration of the exponents'
multiindices. It can be noted how in certain cases the addition of
multiindices (and hence the multiplication of the corresponding monomials)
maps to the addition of their codified representation. For instance:\begin{multline}
c\left(\left[0,0,3\right]\right)+c\left(\left[0,1,0\right]\right)=3+4=7=\\
=c\left(\left[0,1,3\right]\right)=c\left(\left[0,0,3\right]+\left[0,1,0\right]\right),\label{eq:c_01}\end{multline}
where we have noted with $c=c\left(\mathbf{e}\right)$ the function
that codifies a multiindex vector of exponents $\mathbf{e}$. An inspection of
Table \ref{tab:Kronecker-substitution-for} promptly suggests that
for an $m$-variate polynomial up to exponent $n$ in each variable, $c$'s effect
is equivalent to a scalar product between the multiindices vectors
and a constant coding vector $\mathbf{c}$ defined as\begin{equation}
\mathbf{c}=\left[\left(n+1\right)^{m-1},\left(n+1\right)^{m-2},\ldots,n+1,1\right],\end{equation}
so that
\begin{equation}
c\left(\mathbf{e}\right) = \mathbf{c} \cdot \mathbf{e},
\end{equation}
and eq. \eqref{eq:c_01} can be generalised as\begin{equation}
\mathbf{c}\cdot\left(\mathbf{e}_{1}+\mathbf{e}_{2}\right)=\mathbf{c}\cdot\mathbf{e}_{1}+\mathbf{c}\cdot\mathbf{e}_{2}.\label{eq:cod_vector_01}\end{equation}
While this equation is valid in general due to the distributivity
of scalar multiplication, the codification of a multiindex
will produce a unique code only if the multiindex is representable
within the representation defined by $\mathbf{c}$.

This simple example shows how Kronecker substitution constitutes an addition-preserving homomorphism
between the space of integer vectors whose elements are bound in a
finite range and a finite subset of integers. Since the addition of codes maps to the addition of multiindices, the codes can be seen as exponents of a univariate
polynomial.

For use with Laurent polynomials and Poisson series, Kronecker substitution can be conveniently generalised in the following way:

\begin{itemize}
\item we can consider variable codification, i.e., each element of the multiindex
has its own range of variability;
\item we can extend the validity of the codification to negative integers.
\end{itemize}

The first generalisation allows to compact the range of the codes.
If, for instance, the exponent of variable $x$ in the example above
varies only from 0 to 1 (instead of varying from 0 to 3 like for $y$
and $z$), we can avoid codes that we know in advance will be associated
to nonexisting monomials (i.e., all those in which $x$'s exponent
is either 2 or 3). The second generalisation derives from the
fact that under an appropriate extension of the coding vector it is possible to change
$\mathbf{e}_{2}$'s sign in eq. \eqref{eq:cod_vector_01} retaining
the validity of the homomorphism, allowing thus to apply Kronecker substitution
to the multiplication of Poisson series. Poisson series multiplication,
indeed, requires the ability to deal also with negative exponents (since by definition Laurent polynomials may have terms of negative degree);
additionally, the trigonometric multipliers (i.e., the $j$ indices in eq. \eqref{eq:ps_def})
transform under multiplication according to the following elementary trigonometric formulas, which imply
the need to be able to subtract vectors of trigonometric multipliers:\begin{equation}
\begin{gathered}\cos\alpha\cdot \cos\beta=\cos\left(\alpha-\beta\right)+\cos\left(\alpha+\beta\right),\\
\cos\alpha\cdot \sin\beta=\sin\left(\alpha+\beta\right)-\sin\left(\alpha-\beta\right),\\
\sin\alpha\cdot \cos\beta=\sin\left(\alpha-\beta\right)+\sin\left(\alpha+\beta\right),\\
\sin\alpha\cdot \sin\beta=\cos\left(\alpha-\beta\right)-\cos\left(\alpha+\beta\right).\end{gathered}
\label{eq:werner}\end{equation}
Table \ref{tab:Generalised-Kronecker-codification} shows the generalised
Kronecker substitution for a multivariate Laurent polynomial (and Poisson series)
in which the exponents (and trigonometric multipliers) vary on different
ranges, possibly assuming negative values.%
\begin{table*}
\begin{center}
\begin{tabular}{crccc}
\toprule$x_{m-1}$ & $\ldots$ & $x_{1}$ & $x_{0}$ & Code\tabularnewline
\midrule$e_{m-1,\min}$ & $\ldots$ & $e_{1,\min}$ & $e_{0,\min}$ & $0$\tabularnewline
$e_{m-1,\min}$ & $\ldots$ & $e_{1,\min}$ & $1+e_{0,\min}$ & $1$\tabularnewline
$\ldots$ & $\ldots$ & $\ldots$ & $\ldots$ & $\ldots$\tabularnewline
$e_{m-1,\min}$ & $\ldots$ & $e_{1,\min}$ & $e_{0,\max}$ & $e_{0,\max}-e_{0,\min}$\tabularnewline
\midrule$e_{m-1,\min}$ & $\ldots$ & $1+e_{1,\min}$ & $e_{0,\min}$ & $1+e_{0,\max}-e_{0,\min}$\tabularnewline
$e_{m-1,\min}$ & $\ldots$ & $1+e_{1,\min}$ & $1+e_{0,\min}$ & $2+e_{0,\max}-e_{0,\min}$\tabularnewline
$e_{m-1,\min}$ & $\ldots$ & $1+e_{1,\min}$ & $2+e_{0,\min}$ & $3+e_{0,\max}-e_{0,\min}$\tabularnewline
$\ldots$ & $\ldots$ & $\ldots$ & $\ldots$ & $\ldots$\\ \bottomrule\tabularnewline
\end{tabular}
\end{center}

\caption{Generalised Kronecker substitution for an $m$-variate Laurent polynomial
(or Poisson series) in which each exponent (or trigonometric multiplier)
varies on a different range.\label{tab:Generalised-Kronecker-codification}}

\end{table*}
If we define:\begin{eqnarray}
\mathbf{e} & = & \left(e_{0},e_{1},\ldots,e_{m-1}\right),\\
\mathbf{e}_{\min/\max} & = & \left(e_{0,\min/\max},e_{1,\min/\max},\ldots,\right.\nonumber\\
&&\left.e_{m-1,\min/\max}\right),\\
w_{k} & = & 1+e_{k,\max}-e_{k,\min},\\
\mathbf{c} & = & \left(1,w_{0},w_{0}w_{1},w_{0}w_{1}w_{2},\ldots,\vphantom{\Pi_{k=0}^{m-2}w_{k}}\right.\nonumber\\
&&\left.\Pi_{k=0}^{m-2}w_{k}\right),\\
\chi & = & \mathbf{c}\cdot\mathbf{e}_{\min},\end{eqnarray}
it is easy to show that the code of the generic multiindex $\mathbf{e}$
is obtained by\begin{equation}
c\left(\mathbf{e}\right)=\mathbf{c}\cdot\mathbf{e}-\chi.\end{equation}

To recap, this generalisation of the Kronecker substitution technique allows to reduce the multiplication of two multivariate
Poisson series and Laurent polynomials to the multiplication of two univariate polynomials.

\section{The algorithms}
In most applications of practical interest, the univariate polynomials resulting from the application of Kronecker substitution are sparse. E.g., the Fateman benchmark, presented in
\S\ref{sec:benchmarks} and described as a dense benchmark in \cite{monagan09}, after being reduced to a univariate multiplication features roughly $1$ non-null monomial every $300$
in the univariate factors. Series arising in the context of Celestial Mechanics are usually sparser. Because of this, asymptotically fast algorithms for dense multiplication,
such as FFT and Karatsuba, are not usually employed in Celestial Mechanics (see also the discussion in \cite{Fateman03}). The algorithms described in this paper thus employ
schoolbook (aka ordinary) multiplication: each monomial of the first univariate polynomial factor is multiplied by all the monomials of the second factor.

Desirable properties of a multiplication algorithm for Celestial Mechanics applications include:
\begin{itemize}
 \item the ability to operate on multiple types of numerical coefficients (e.g., reals, rationals, integers, both in machine precision and multiprecision, complex numbers, intervals);
 \item the ability to efficiently truncate multiplication.
\end{itemize}
The second requirement stems from the observation that the number of terms of the series involved in many practical calculations tends to explode during multiplication,
if not controlled properly.
Typical truncation criterions concern quantities such as the absolute value of the numerical coefficients, the minimum exponents of one or more polynomial variables and
the order of the Fourier harmonics (see the discussion in \cite{mcm}, Chapter 2, Section 3).

In any case, in order to truncate efficiently (i.e., without having
to compute a term to discard it afterwards), a truncation criterion may define an ordering over the terms of the series being multiplied. This way, while multiplying
one monomial of the first factor and iterating over the monomials of the second factor, it may be possible to skip all monomial-by-monomial multiplications from a certain point onwards.
The multiplication algorithm, hence, should ideally be flexible enough to allow for this kind of truncation methodology without
a negative impact on performance.

\subsection{Moderately-sparse multiplication}
The first algorithm we present is very simple and suitable for moderately-sparse multiplication.
The two univariate polynomial factors are represented
as vectors of coefficient-exponent pairs (a representation often referred to as \emph{sparse distributed}). The resulting univariate polynomial
is represented instead as a vector of coefficients, with the positional index of each coefficient implicitly encoding the corresponding exponent
(a \emph{dense distributed} representation). This vector of coefficients is initialised with null values.

\begin{figure}[ht]
\begin{center}
\includegraphics[width=7cm]{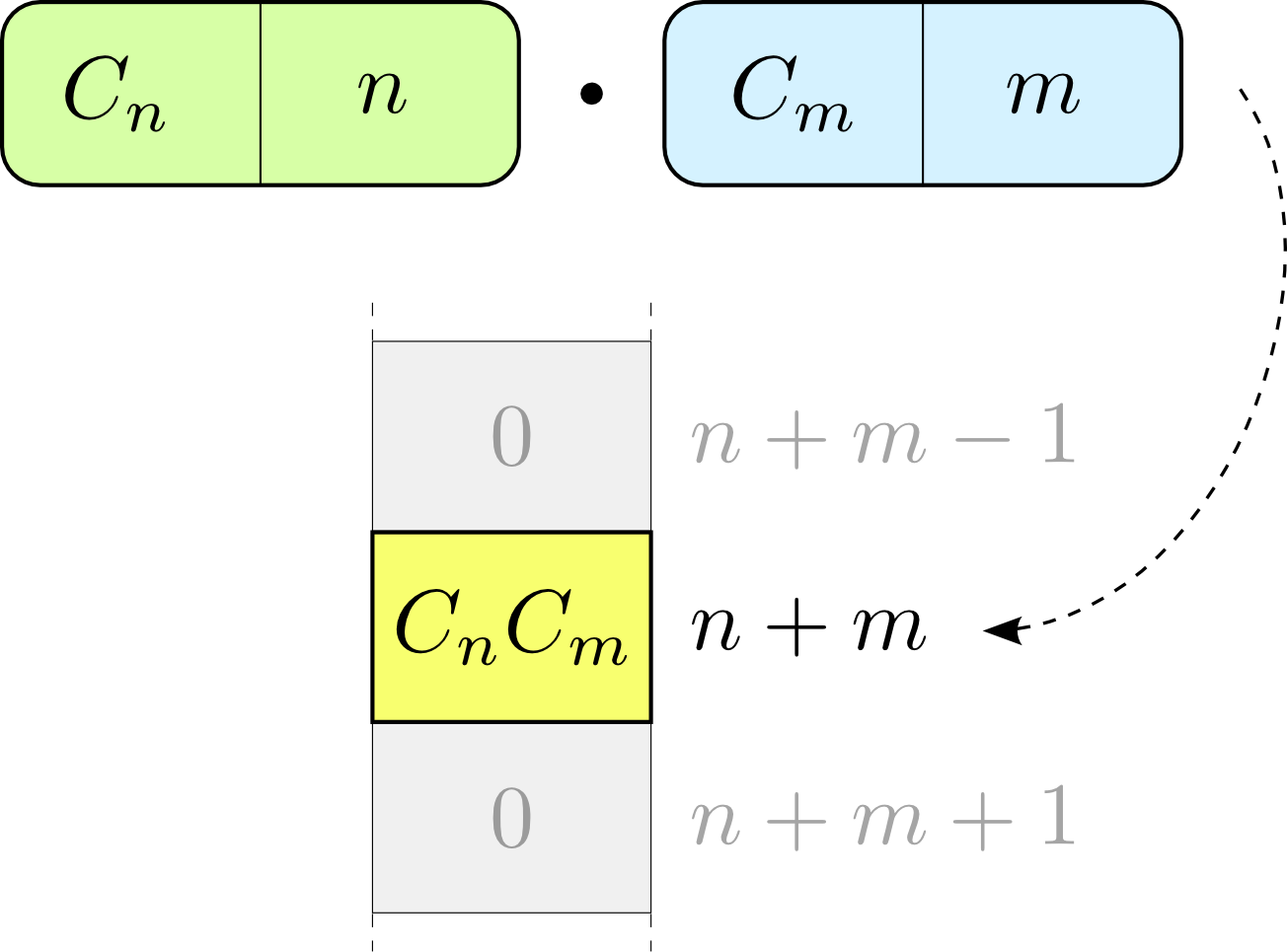}
\caption{Moderately-sparse multiplication: the multiplication of two terms $\left(C_n,n\right)$ and $\left(C_m,m\right)$ of the univariate polynomial factors
(where $n$ and $m$ are the exponents to which coefficients $C_n$ and $C_m$ are associated) produces coefficient
$C_nC_m$, which is stored in the output array of coefficients at position $n+m$.}
\label{fig:mod_sparse}
\end{center}
\end{figure}

Each monomial-by-monomial multiplication generates a coefficient, which is added to
the coefficient in the output vector at the positional index equal to the sum of the exponents of the monomial factors (see
Figure \ref{fig:mod_sparse}). In case of Poisson series multiplication, the same coefficient is also added to the coefficient at the positional index
equal to the subtraction of the exponents of the monomial factors (as per eqs. \eqref{eq:werner}). The exponents of the
monomials resulting from the multiplication, in other words, are used as trivial (and perfect)
hash values for accumulation in a lookup array of coefficients.

The practical performance of this algorithm crucially depends on two optimisations related to cache memory:
\begin{itemize}
 \item \textbf{cache blocking}: the two polynomial factors are subdivided logically into blocks, and, rather than iterating over
 all the monomials of the second factor having fixed a monomial in the first one (in a doubly-nested \texttt{for} loop fashion),
 multiplication is performed block-by-block. The
 effect is the same as if the two factors were subdivided into smaller polynomials to be multiplied separately;
 \item \textbf{monomial ordering}: the two polynomial factors are sorted in ascending order according to the degree of the monomials.
\end{itemize}
The combined effect of these optimisations is a cache-friendly memory access pattern which promotes temporal and spatial locality
of reference through:
\begin{enumerate}
 \item sequential writes into the output coefficient vector, by virtue of the monomial ordering,
 \item short-term reuse of data already present into the cache memory, by virtue of cache blocking.
\end{enumerate}
\begin{figure*}[ht]
\begin{center}
\includegraphics[width=8.5cm]{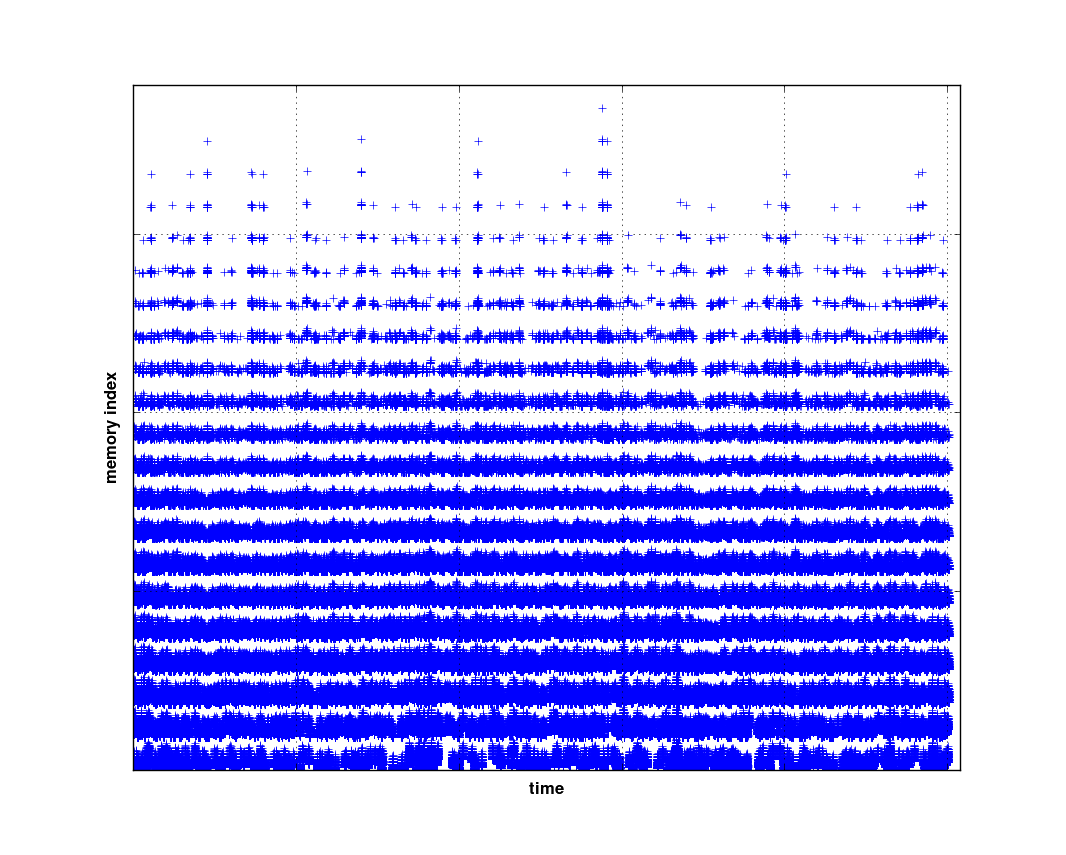}
\includegraphics[width=8.5cm]{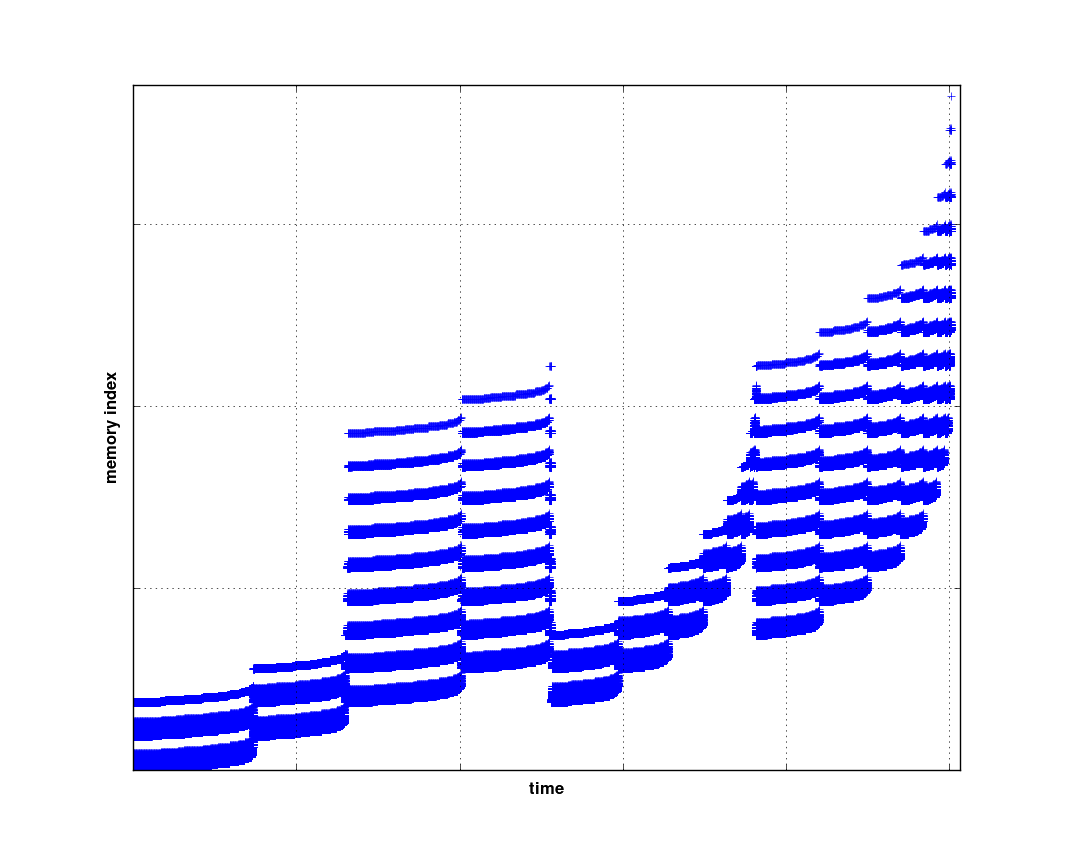}
\caption{Typical memory access patterns for moderately-sparse multiplication, with (right) and without (left) cache-blocking and term ordering optimisation. The $x$ axis
displays the time at which the memory write is performed in the output coefficient vector, the $y$ axis displays its location within the vector.}
\label{fig:mem_pattern_mod}
\end{center}
\end{figure*}
A typical memory access pattern for moderately-sparse multiplication is displayed in Figure \ref{fig:mem_pattern_mod}.

These optimisations cope well also when a different monomial ordering is required by the truncation criterion. Indeed,
to retain much of the cache-friendly memory access pattern, it is enough to first order the monomials according to the degree and then
reorder within each block the monomials according to the order requested by the truncation criterion.

\subsection{Highly-sparse multiplication}
In case of highly-sparse multiplication, the algorithm described above may not be applicable for the following reasons:
\begin{itemize}
 \item the output coefficient array would occupy much too memory storage,
 \item the high sparsity would result in a large array whose initialisation overhead would outweigh the time spent in the actual multiplication.
\end{itemize}
Such occurrences are detectable through an analysis of the densities of the univariate polynomial factors.

The algorithm we have adopted to cope with the highly-sparse case is based on hashing techniques, and it essentially replaces the output coefficient array of the first algorithm
with a cache-friendly hash table. Our design is just one possibility of cache-friendly hash table: other designs, including cuckoo hashing \cite{cuckoo_hash} and hopscotch hashing
\cite{hopscotch_hash} (but also linear probing \cite{knuth98_3}) may be effective too.

\begin{figure*}[ht]
\begin{center}
\includegraphics[width=12cm]{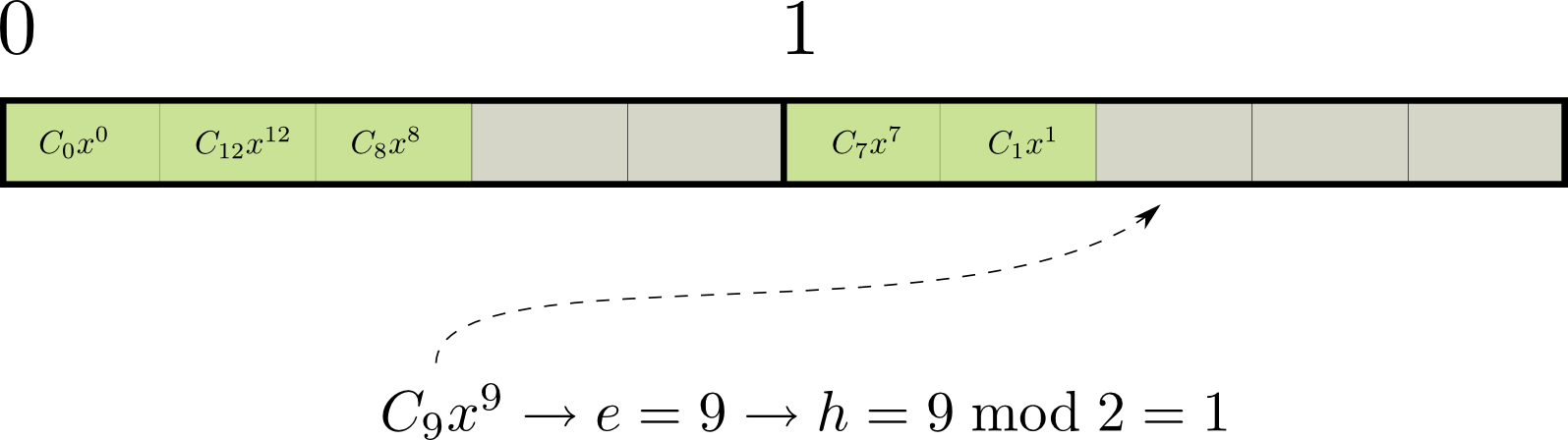}
\caption{Insertion of monomial $C_9x^9$ into a partially-filled hash table with $n=10$, $N=2$ and $m=5$.}
\label{fig:hash_table}
\end{center}
\end{figure*}

The hash table is implemented as a contiguous memory area of size $n$ logically subdivided into $N$ adjacent buckets of maximum size $m = n / N$. The buckets store
coefficient-exponent pairs (i.e., in a sparse distributed representation). The exponents of the monomials resulting from the multiplication of the univariate polynomial factors are used,
after modulo reduction, as hash values to accumulate the monomials into the hash table. E.g., when a monomial with exponent $e$ is produced during multiplication, its hash value $h$
is computed,
\begin{equation}
 h = e \bmod{N},
\end{equation}
and the monomial is inserted into the $h$-th bucket (see Figure \ref{fig:hash_table}). If the bucket already contains a term with the same exponent $e$, then the coefficient of the incoming monomial is added to the existing one.
Otherwise the monomial is appended at the end of the bucket. Since the maximum size of the buckets is $m$, insertion of a new monomial can fail when a bucket already contains $m$ elements;
in this case the size of the table is increased and the elements re-hashed.

In order to reduce the need for resizing, an additional ``overflow'' bucket is allocated. When an insertion fails, the monomial is inserted into the overflow bucket, and the
hash table resize is delayed until the number of monomials in the overflow bucket reaches a certain threshold $s$. The overflow bucket must also be checked whenever, during a probe of the table,
a full bucket is encountered.

\begin{figure}[ht]
\begin{center}
\includegraphics[width=8.4cm]{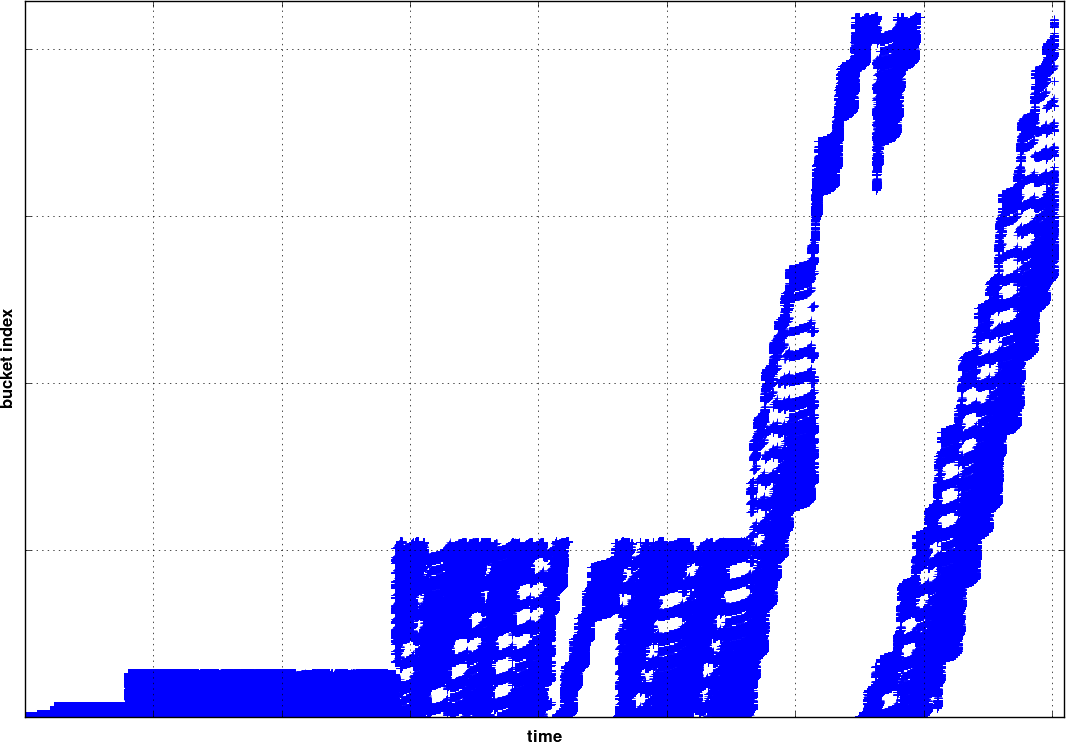}
\caption{Typical memory access pattern for highly sparse multiplication, with cache blocking and term ordering optimisation. The $x$ axis
displays the time at which the memory write is performed in the hash table, the $y$ axis displays the index of the bucket.}
\label{fig:pattern_hash}
\end{center}
\end{figure}

The cache memory optimisations introduced for the moderately sparse case can be used also for highly sparse multiplication. The only modification concerns term ordering: instead
of sorting the univariate polynomial factors according to the exponents, the terms are sorted according to the exponents modulo $N$. Elementary properties of modular arithmetics
ensure then that consecutive write operations happen on consecutive buckets (see Figure \ref{fig:pattern_hash}). Each time a re-hash operation takes place, the polynomial factors
must be re-ordered.

\section{Parallelisation}
The first algorithm has been parallelised for shared memory multicore architectures using multiple threads of execution. The parallelised algorithm assigns to each thread a portion
of the univariate polynomial factors, and all threads concurrently write the results into the same output coefficient array. The algorithm relies on the cache memory optimisations
described earlier to avoid contention. The following example explains how.

Let us suppose that the univariate polynomial factors, $P_1$ and $P_2$ have been divided into blocks. Since the monomials have been sorted according to their exponent, the $n$-th block
of a polynomial will contain monomials whose exponents are bound in an interval
\begin{equation}
 I_n = \left[ e_{n,i},  e_{n,f}\right],
\end{equation}
whereas the block immediately following will feature exponents in the interval
\begin{equation}
 I_{n+1} = \left[ e_{n+1,i},  e_{n+1,f}\right],
\end{equation}
with
\begin{equation}
 e_{n,f} < e_{n+1,i}.\label{eq:int_ineq}
\end{equation}
From elementary interval arithmetics it follows that multiplying $P_1$'s block $I_{n}^{(1)}$ by $P_2$'s block $I_{n}^{(2)}$
will produce monomials whose exponents will be in the range
\begin{equation}
\left[ e_{n,i}^{(1)} + e_{n,i}^{(2)} ,  e_{n,f}^{(1)} + e_{n,f}^{(2)} \right].
\end{equation}
Similarly, multiplying $P_1$'s block $I_{n+1}^{(1)}$ by $P_2$'s block $I_{n+1}^{(2)}$
will produce monomials whose exponents will be in the range
\begin{equation}
\left[ e_{n+1,i}^{(1)} + e_{n+1,i}^{(2)} ,  e_{n+1,f}^{(1)} + e_{n+1,f}^{(2)} \right].
\end{equation}
But then, from eq. \eqref{eq:int_ineq}, the following inequality holds:
\begin{equation}
 e_{n,f}^{(1)} + e_{n,f}^{(2)} < e_{n+1,i}^{(1)} + e_{n+1,i}^{(2)}.
\end{equation}
This implies that the exponents resulting from the multiplication of $I_{n}^{(1)}$ by $I_{n}^{(2)}$ and those of $I_{n+1}^{(1)}$ by $I_{n+1}^{(2)}$ will not overlap.
Hence, since the exponents encode also the memory position in the output coefficient array, this also means that performing the multiplications concurrently will not
cause any contention issue, since the interested memory areas are disjoint.

The parallelised algorithm can now be explained by an example. Let us suppose, for the sake of simplicity, that there are $4$ available threads,
that both univariate factors have the same length and that the chosen
block size divides this length exactly, so that each factor is consituted of $n > 4$ blocks. The first $4$ blocks of the first factor are assigned one per thread (so that thread $1$ is assigned
block $1$, thread $2$ block $2$ and so on). Each thread starts by multiplying concurrently its block by the corresponding block in the second factor (block $1$ by block $1$, $2$ by $2$, etc.).
A condition variable is used as a synchronisation barrier which all threads must reach after they have completed their block-by-block
multiplications. At this point, each thread moves to the next block in the second factor, so that at this point thread $1$ is multiplying block $1$ of the first
factor by block $2$ of the second factor, thread $2$ is performing $2$ by $3$ and so on. The same pattern repeats until the last block of the second factor is reached, where each thread alternately
sleeps for $3$ ($=\textnormal{number of threads} -1$) iterations. These ``silent'' iterations are needed in order to make sure that concurrent operations are performed on consecutive blocks, as explained above.
At this point the first four blocks of the first factor have been multiplied by all the blocks of the second factor;
the threads acquire the next four blocks in the first factor and restart the same procedure from the top of the second factor.
The sequence of block-by-block multiplications is summarised in Table \ref{tab:block_parallel}.

This same algorithm is also applicable, albeit with some additional complications, in case of highly-sparse multiplication, where the ordering of the term according to the exponent modulo the hash table's
size ensures concurrent write on disjoint memory areas (apart from writes into the overflow bucket, which must be protected by locking). We are currently in the process of implementing this algorithm
for the highly-sparse case.

\begin{table*}
\begin{center}
\begin{tabular}{ccc|c|c|c|c|c|c|c|c|c|c}
\toprule
$P_1$ & &\multicolumn{11}{c}{$P_2$} \\
\midrule
$\mathbf{1}$ & $\rightarrow$ &$1$ & $2$ & $3$ & $\cdots$ & $n-3$ & $n-2$ & $n-1$ & $n$ & -- & -- & -- \\
$\mathbf{2}$ & $\rightarrow$ &$2$ & $3$ & $4$ & $\cdots$ & $n-2$ & $n-1$ & $n$ & -- & -- & -- & $1$ \\
$\mathbf{3}$ & $\rightarrow$ &$3$ & $4$ & $5$ & $\cdots$ & $n-1$ & $n$ & -- & -- & -- & $1$ & $2$ \\
$\mathbf{4}$ & $\rightarrow$ &$4$ & $5$ & $6$ & $\cdots$ & $n$ & -- & -- & -- & $1$ & $2$ & $3$ \\
\bottomrule
\end{tabular}
\end{center}
\caption{Parallel multiplication of the first four blocks of the first factor $P_1$ by all the blocks of the second factor $P_2$. The ``--'' symbols represent silent iterations during which no work is performed
by the thread, and the vertical lines represent thread barriers. Concurrent multiplications follow column-by-column.}
\label{tab:block_parallel}
\end{table*}

\section{Benchmarks}
\label{sec:benchmarks}
The algorithms described in this paper have been implemented in C++ within a specialised algebraic manipulator for Celestial Mechanics called Piranha \cite{piranha}. For comparison,
we quote the results from \cite{monagan09}, where the tested programs are SDMP \cite{monagan09}, TRIP \cite{gastineau2006}, PARI \cite{PARI}, Maple \cite{Maple} and Magma \cite{bosma97}.
Here we limit the comparison to SDMP, which provides the best performance according to \cite{monagan09} (other timings can be found in \cite{monagan09}). The three benchmarks we present are:
\begin{itemize}
 \item Fateman's benchmark \cite{Fateman03}: calculate $f \cdot g$, where $f=\left( 1+x+y+z+t\right)^{30}$ and $g = f+1$. $f$ and $g$ consist of 46376 terms. This benchmark
 is suitable for the first algorithm.
 \item ELP Poisson series multiplication: calculate $\textnormal{ELP3}^3 \cdot \textnormal{ELP3}^3$, where ELP3 is the Poisson series representing the lunar distance
 in the main problem of the Lunar Theory ELP2000 \cite{ELP2000}. ELP3 is a Poisson series with numerical coefficients and $\textnormal{ELP3}^3$ consists of 60204 terms. This benchmark
 is suitable for the first algorithm.
 \item Monagan-Pearce sparse (MP-sparse) benchmark \cite{monagan09}: calculate $f\cdot g$, where
\[
f = \left( 1+x+y+2z^2+3t^3+5u^5 \right)^{12}
\]
and
\[
g = \left( 1+u+t+2z^2+3y^3+5x^5 \right)^{12}.
\]
 $f$ and $g$ consist of 6188 terms. This benchmark is suitable for the second algorithm.
\end{itemize}
The hardware and software configurations on which the tests were performed are displayed in Table \ref{tab:arches}. Regarding our Corei7 configuration, we have to remark here that we gained access
to the machine shortly before submitting the paper and thus we were not able to conduct extensive testing. We will produce more complete benchmarks on this architecture in the future.
\begin{table*}
\begin{center}
\begin{tabular}{lll}
\toprule
\textbf{Short name} & \textbf{Hardware} & \textbf{Software} \\
\midrule
\multirow{2}{*}{Core2Quad} & Intel Core2 Q6600, 2.4GHz,& Linux 2.6.31.5, 64 bit,\\
& 4 cores, 2 x 4MB L2 cache, 4GB DDR2& GCC 4.4.2, GMP 4.3.2 \\
\multirow{2}{*}{Core2Duo} & Intel Core2 T7100, 1.8GHz, & Linux 2.6.31.2, 64-bit, \\
& 2 cores, 2MB L2 cache, 2GB DDR2 & GCC 4.4.2, GMP 4.3.1\\
\multirow{2}{*}{PPC64} & 2 x IBM PowerPC 970, 2GHz, & Linux 2.6.32, 64-bit\\
& 2 cores, 512KB L2 cache, 8GB DDR2 & GCC 4.4.2, GMP 4.3.1 \\
\multirow{2}{*}{Atom} & Intel Atom N270, 1.6GHz, & Linux 2.6.31.1, 32-bit,\\
&2 cores (Hyper-threading), 512KB L2 Cache, 1GB DDR2& GCC 4.4.2, GMP 4.3.1 \\
\multirow{2}{*}{Xeon} & 2 x Intel Xeon X5355, 2.66GHz,& Linux 2.6.28, 64-bit,\\
& 8 cores, 2 x 4MB L2 cache, 8GB DDR2 & GCC 4.3.3, GMP 4.2.4\\
\multirow{2}{*}{Corei7} & Intel Core i7-940, 2.93GHz, 8 cores (Hyper-threading),& OSX 10.6, 64-bit,\\
& 4 x 256KB L2 cache, 8MB L3 cache, 4GB DDR3 & Xcode 3.2, GMP 4.3.2\\
\multirow{2}{*}{SDMP-Core2} & Intel Core2 Q6600, 2.4GHz,& Linux 2.6.26, 64-bit, \\
& 4 cores, 2 x 4MB L2 cache, 4GB DDR2 & GCC 4.3.2, GMP 4.2.2 \\
\multirow{2}{*}{SDMP-Corei7} & Intel Core i7-920, 2.66GHz, 4 cores & Linux 2.6.27, 64-bit,\\
& 4 x 256KB L2 cache, 8MB L3 cache, 6GB DDR3 & GCC 4.3.2, GMP 4.2.2 \\
\bottomrule
\end{tabular}
\end{center}
\caption{Hardware and software configurations used in the benchmarks. The SDMP configurations are quoted from \cite{monagan09}.}
\label{tab:arches}
\end{table*}

In the results we report both the wall clock timings and the clock cycles per monomial-by-monomial multiplication -- shortened as ``ccpm''. An ``optimal'' algorithm should be able
to have a count of ccpm close to the number of clock cycles needed to multiply and add coefficients on a specific architecture, hence minimising the bookkeeping overhead and
cache misses.

\subsection{Sequential benchmarks}
\begin{table}
\begin{center}
\footnotesize
\begin{tabular}{lllrr}
\toprule
\textbf{Test} & \textbf{Coefficient} & \textbf{System} & \textbf{Time} & \textbf{ccpm} \\
\midrule
Fateman & double & Core2Quad & 4.29s & 4.8\\
Fateman & double & Core2Duo & 5.62s & 4.6\\
Fateman & double & PPC64 & 4.96s & 4.6\\
Fateman & double & Xeon & 3.73s & 4.6\\
Fateman & double & Atom & 20.15s & 15.0\\
Fateman & GMP mpz & Core2Quad & 67.90s & 75.8\\
Fateman & 61-bit integer & SDMP-Core2 & 60.25s & 67.2\\
Fateman & 61-bit integer & SDMP-Corei7 & 70.59s & 85.3\\
ELP & double & Core2Quad & 15.62s & 10.3\\
MP-sparse & double & Core2Quad & 1.71s & 107.2 \\
MP-sparse & double & Xeon & 1.59s & 110.5 \\
MP-sparse & double & Corei7 & 1.15s & 88.0 \\
MP-sparse & 37-bit integer & SDMP-Core2 & 1.86s & 116.6 \\
MP-sparse & 37-bit integer & SDMP-Corei7 & 1.56s & 108.4 \\
\bottomrule
\end{tabular}
\normalsize
\end{center}
\caption{Serial benchmarks. The SDMP results are quoted from \cite{monagan09}.}
\label{tab:ser_benchmarks}
\end{table}

The results of the benchmarks in single-thread mode are summarised in Table \ref{tab:ser_benchmarks}. Some considerations:
\begin{itemize}
 \item we have performed the Fateman benchmark using both double precision and multiprecision integer coefficients (the latter implemented using the GMP library \cite{GMP}).
 For the same test, SDMP is using 61-bit integer coefficients as input and 128-bit integer coefficients as output.
 \item The first algorithm is able to deliver close to optimal performance in the Fateman benchmarks. Indeed, according to \cite{fog_instrtables}, floating-point multiplication
 on most Intel CPUs has a latency of $4-5$ clock cycles. The performance degradation on the Atom might be explained by the fact the Atom is the only CPU, among those tested, that operates in-order
 (thus resulting in less optimisations available to the compiler).
 \item By comparison, Roman Pearce reported to us in a personal communication that coefficient multiplication by SDMP in the Fateman benchmark costs around 18 clock cycles.
 \item Performance in the ELP benchmark is also close to optimal, considering that Poisson series multiplication is slightly more complicated than polynomial multiplication.
 \item In the highly sparse MP-sparse benchmark, there is much more algorithmic overhead and effective cache memory usage is more difficult to achieve than in the Fateman benchmark.
 Our algorithm seems to benefit greatly from the high amount of cache and the DDR3 memory available on the Corei7 configuration.
\end{itemize}

\subsection{Parallel benchmarks}

\begin{figure}
\begin{center}
\includegraphics[width=8.4cm]{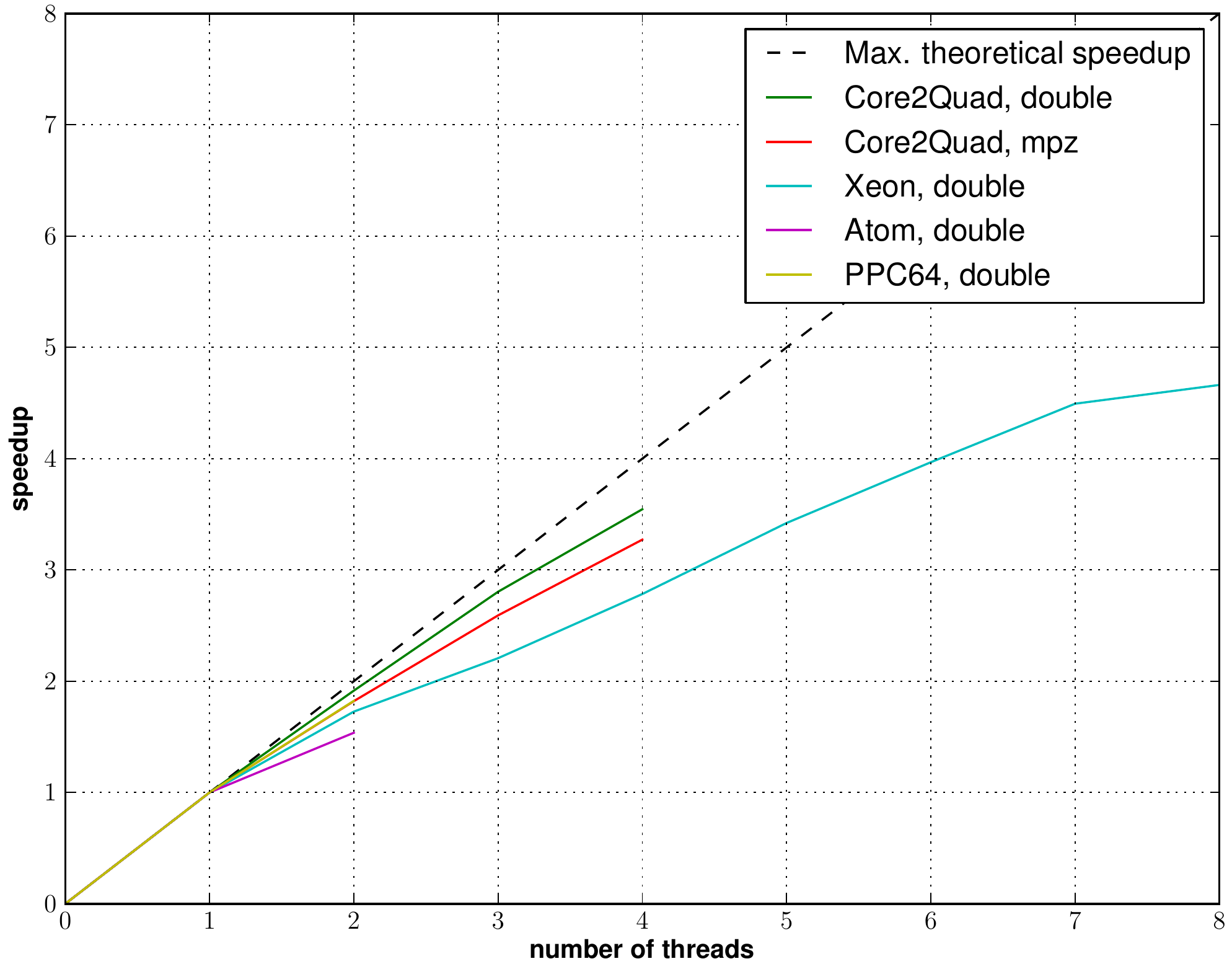}
\caption{Speedup for the parallel Fateman benchmark on various systems. The data points for PPC64 overlap those for Core2Quad, mpz.}
\label{fig:par_bench}
\end{center}
\end{figure}

The speedups resulting from the parallel Fateman benchmark on various configurations are displayed in Figure \ref{fig:par_bench}.
The speedups, calculated with respect to the sequential algorithm, are almost linear for Core2Quad, both with double precision and mpz coefficients, and also for PPC64.
Also on the Atom a $\sim1.5$ speedup is measured, despite the use of hyper-threading.

The oddball is clearly the Xeon system, which, despite exhibiting performance improvements up to all the 8 cores, features a smaller speedup with respect
to the other quad-core processors benchmarked. Moreover, the results of multiple benchmark runs on this configuration resulted in timings proportionally
more unstable than on the other systems. A possible explanation for this behaviour is that the machine is shared among multiple users, and, despite we made sure
that no other onerous computations were going on as we performed the tests, other unrelated process (daemons, running graphical sessions, etc.)  may have interfered
with the benchmark. Another point of interest is that the Xeon system and the PPC64 systems are the only two configurations featuring two physically separated processors
(as opposed to the other single-processor multi-core configurations). It may be possible that inter-processor communication consitutes a bottleneck for the algorithms presented here.

\section{Conclusions}
In this paper we have presented two algorithms for the multiplication of sparse Laurent polynomials and Poisson series on modern hardware architectures. The benchmarks performed
on various hardware and software configurations suggest that these algorithms are competitive, performance-wise, with the fastest algorithm currently known and implemented in the SDMP
library.

Future work will focus on the parallelisation of the highly-sparse algorithm. We also need to test the algorithms on architectures with a higher number of cores, in order to verify
where the speedup limit lies; further tests are also needed to better assess the effectiveness of the algorithms on architectures with shared L3 cache,
such as the Intel Core i7. Finally, we need to investigate the suboptimal speedup exhibited by the Xeon system.

\section{Acknowledgements}
We are very grateful to Roman Pearce for helpful discussions and useful comments.

\bibliographystyle{abbrv}
\bibliography{issac2010_paper}

\end{document}